\documentclass[%
 reprint,
 superscriptaddress,
 a4paper,
 showpacs,
 showkeys,
 nofootinbib,
 amsmath,amssymb,
 aps,
10.0pt]{revtex4-1}

\usepackage{graphicx}
\usepackage{float}
\usepackage[utf8]{inputenc}
\usepackage{bm}
\usepackage[colorlinks=true, linkcolor=blue,citecolor=blue,allcolors=blue]{hyperref}
\usepackage{tabularx}
\usepackage{xcolor}
\usepackage[capitalize]{cleveref}
\usepackage{makecell}
\usepackage{cancel}
\usepackage{xspace}
\usepackage{parskip}
\usepackage{xcolor,colortbl}
\usepackage{slashed}

\newcommand{\fmm}{${\rm fm^{-3}}$\xspace}
\newcommand{\fm}{${\rm fm^{3}}$\xspace}

\graphicspath{{./graphics/}}

\begin{document}

\title{Repulsive properties of hadrons in lattice QCD data and neutron stars}
\date{\today}
\author{Anton~Motornenko}
\affiliation{
Institut f\"ur Theoretische Physik,
Goethe Universit\"at, D-60438 Frankfurt am Main, Germany}
\affiliation{
Frankfurt Institute for Advanced Studies, Giersch Science Center,
D-60438 Frankfurt am Main, Germany}
\author{Somenath~Pal}
\affiliation{
Department of Physics,
University of Calcutta,
92, A.P.C. Road, Kolkata-700009, India}
\author{Abhijit~Bhattacharyya}
\affiliation{
Department of Physics,
University of Calcutta,
92, A.P.C. Road, Kolkata-700009, India}
\author{Jan~Steinheimer}
\affiliation{
Frankfurt Institute for Advanced Studies, Giersch Science Center,
D-60438 Frankfurt am Main, Germany}
\author{Horst~Stoecker}
\affiliation{
Institut f\"ur Theoretische Physik,
Goethe Universit\"at, D-60438 Frankfurt am Main, Germany}
\affiliation{
Frankfurt Institute for Advanced Studies, Giersch Science Center,
D-60438 Frankfurt am Main, Germany}
\affiliation{
GSI Helmholtzzentrum f\"ur Schwerionenforschung GmbH, D-64291 Darmstadt, Germany}

\begin{abstract}
Second-order susceptibilities $\chi^{11}_{ij}$ of baryon, electric, and strangeness, $B$, $Q$, and $S$, charges, are calculated in the Chiral Mean Field (CMF) model and compared to available lattice QCD data. The susceptibilities are sensitive to the short range repulsive interactions between different hadron species, especially to the hardcore repulsion of hyperons. Decreasing the hyperons size, as compared to the size of the non-strange baryons, does improve significantly the agreement of the CMF model results with the Lattice QCD data. The electric charge-dependent susceptibilities are sensitive to the short range repulsive volume of mesons. The comparison with lattice QCD data suggests that strange baryons, non-strange mesons and strange mesons have significantly smaller excluded volumes than non-strange baryons. 
The CMF model with these modified hadron volumes allows for a mainly hadronic description of the QCD susceptibilities significantly above the chiral pseudo-critical temperature.
 This improved CMF model which is based on the lattice QCD data, has been used to study the properties of both cold QCD matter and neutron star matter. 
The phase structure in both cases is essentially unchanged, i.e. a chiral first-order phase transition occurs at low temperatures ($T_{\rm CP}\approx 17$ MeV), and hyperons survive deconfinement to higher densities than non-strange hadrons. The neutron star maximal mass remains close to 2.1$M_\odot$ and the mass-radius diagram is only modified slightly due to the appearance of hyperons and is in agreement with astrophysical observations.
\end{abstract}

\maketitle

\section{Introduction}

	The study of the properties of hot and dense strongly interacting matter is an active area of fundamental research. For some decades now the focus is on the phase structure of the matter created in ultra relativistic heavy-ion collisions and dense compressed matter in neutron stars (NS). Experimental facilities as the Large Hadron Collider (LHC) at CERN and the Relativistic Heavy Ion Collider (RHIC) at BNL have provided a lot of information about the hot and dense strongly interacting matter. Upcoming experimental programs like the Facility for Anti-Proton and Ion Research (FAIR) and  the Nuclotron-based Ion Collider fAcility (NICA) will yield a better understanding of the properties of such matter. In spite of extensive theoretical and experimental research, the current understanding of the phase structure of strongly interacting matter is far from complete. Lattice QCD calculations show that, at high temperatures and vanishing chemical potential, both the chiral transition and the transition from confined hadronic matter to deconfined quark matter occurs as a crossover~\cite{Borsanyi:2013bia,Bazavov:2014pvz}.
	On the other hand, at low temperatures and high chemical potentials, effective model studies suggest that a First-order phase transition~\cite{Halasz:1998qr,Fukushima:2010bq} could be realized. This suggests the existence of a Critical End Point (CEP) in the phase diagram of QCD.
	The extension of lattice QCD to finite chemical potentials often relies on a Taylor expansion with the expansion coefficients being related to the conserved charge susceptibilities~\cite{Allton:2002zi}. Alternatively, a continuation from imaginary chemical potentials is employed to extrapolate to high $\mu_B$~\cite{deForcrand:2002hgr,Vovchenko:2017xad,Vovchenko:2017gkg}. These methods are needed as the fermion sign problem prohibits direct calculations at high finite baryon chemical potentials. The QCD thermodynamics at vanishing chemical potentials is connected with the phase structure at finite chemical potentials. Hence, these susceptibilities serve as important indicators of the validity of effective models and for the effective (quasi-)particle interactions employed in these models. In particular, the susceptibilities have proven to be sensitive to the density dependent repulsive interactions of quarks (see e.g. \cite{Steinheimer:2010sp,Steinheimer:2014kka}). 
	QCD gives rise to a rich spectrum of hadronic states with an even richer set of reciprocal interactions. Though attempts have been made to extract the properties of these interactions directly from scattering data \cite{Dashen:1969ep,Venugopalan:1992hy,Lo:2017sde,Vovchenko:2017drx}, calculations of the thermodynamic properties of strongly interacting hadronic matter are mostly based on phenomenological considerations. In this paper we  show how the second-order conserved charge susceptibilities, as calculated on the lattice, can be used to extract the features of the different hadronic repulsive interactions. Furthermore, these calculations are used to understand the role of these interactions for the structure of the QCD phase diagram and for neutron star properties.
    	
	The paper is organized as follows: the next section,~\ref{sec:CMF_def}, discusses, in some detail the  Chiral Mean Field model. Section~\ref{sec:latice-analysis} compares the CMF results on fluctuations and on correlations with the lattice data. Section~\ref{sec:phase} is devoted to study the QCD-based CMF-phase structure at low and high densities. Section~\ref{sec:NS} presents results for the $\beta$-equilibrated ultra-high density matter and the structure of neutron stars. Section~\ref{sec:summary} summarizes our results.

\section{Chiral Mean Field model}
\label{sec:CMF_def}

Many effective models have been employed to describe and interpret the lattice QCD data (see e.g. \cite{Ratti:2005jh,Roessner:2006xn,Fukushima:2003fw,Schaefer:2007pw,Luecker:2013oda}), often based on quark quasiparticle or functional renormalization group methods. These models are mainly concentrated on the description of QCD properties at $\mu_B=0$ on the level of partonic degrees of freedom. For these models the finite baryonic densities are not of interest, and the role of the confined phase and hadronic interactions cannot be estimated there.
Attempts to extend lattice-QCD-inspired approaches to finite densities have been made before, mostly based on quark-gluon quasi particle models \cite{Peshier:1999ww,Ivanov:2005be,Khvorostukhin:2010aj}. In the following, we describe a related ansatz which attempts to simultaneously describe the hadronic and deconfined degrees of freedom in a self-consistent approach. The proposed ansatz, the CMF model, does not require any phase matching and thus allows for a continuous transition from confined to deconfined degrees of freedom while at the same time giving a proper description of nuclear matter and neutron star matter phenomenology.
To be able to make conclusive statements for high-density QCD it is necessary to also study the susceptibilities of conserved charges as done in this work, in addition to having a proper description of hadronic and nuclear matter. 

The Chiral ${\rm SU}(3)$-flavor parity-doublet Polyakov-loop quark-hadron mean-field model (or the CMF model) describes the thermodynamics of strongly interacting matter on both the hadronic and quark level. The CMF model allows us to calculate the equation of state (EOS) of QCD matter at wide range of temperatures and densities. It incorporates major concepts of QCD phenomenology: chiral interactions in the baryon octet~\cite{Papazoglou:1998vr}, the full PDG hadron list~\cite{Tanabashi:2018oca}, excluded volume repulsive interactions among all hadrons~\cite{Rischke:1991ke, Steinheimer:2010ib}, baryon parity doubling~\cite{Detar:1988kn}, and quarks coupled to an effective Polyakov loop potential (similar to the Polyakov Nambu Jona-Lasinio model~\cite{Fukushima:2003fw}).

The main component of the CMF model is the three flavor chiral Lagrangian for strange hadronic matter first introduced in Ref.~\cite{Papazoglou:1998vr}. The Lagrangian ${\cal L}_{{\rm SU}(3)_f}$ consists of the following parts:
\begin{eqnarray}
{\cal L}_{SU(3)_f}={\cal L_B}  + U_{\rm sc} + U_{\rm vec}
\end{eqnarray}
where ${\cal L_B}$ describes scalar and vector mean-field interactions among the ground-state octet baryons and their parity partners:
\begin{eqnarray}
{\cal L_B} &=& \sum_b (\bar{B_b} i \slashed{\partial} B_b)
+ \sum_b  \left(\bar{B_b} m^*_b B_b \right) \nonumber \\ &+&
\sum_b  \left(\bar{B_b} \gamma_\mu (g_{\omega b} \omega^\mu +
g_{\rho b} \rho^\mu + g_{\phi b} \phi^\mu) B_b \right) \,,
\label{lagrangian2}
\end{eqnarray}
where the index $b$ runs through all ground-state baryons, $p$, $n$, $\Lambda$, $\Sigma^{+,0,-}$, $\Xi^{0,-}$,  and their respective parity partners, $N(1535)^{+,0}$, $\Lambda(1405)$, $\Sigma(1750)^{+,0,-}$, $\Xi(1950)^{0,-}$.
$U_{\rm sc}$ describes the potential of the scalar $\sigma$ and $\zeta$ fields, and $U_{\rm vec}$ is the potential of the vector $\omega$, $\rho$, and $\phi$ fields.

The effective masses of the ground-state  octet baryons and their parity partners (assuming isospin symmetry) read~\cite{Steinheimer:2011ea}:
\begin{eqnarray}
m^*_{b\pm} &=& \sqrt{ \left[ (g^{(1)}_{\sigma b} \sigma + g^{(1)}_{\zeta b}  \zeta )^2 + (m_0+n_s m_s)^2 \right]} \nonumber \\ 
& \pm & g^{(2)}_{\sigma b} \sigma \ ,
\end{eqnarray}
where the various coupling constants $g^{(*)}_{*b}$ are determined by vacuum masses and by nuclear matter properties. $m_0$ refers to a bare mass term of the baryons which is not generated by the breaking of chiral symmetry, and $n_s m_s$ is the ${\rm SU}(3)_f$-breaking mass term that generates an explicit mass corresponding to the strangeness $n_s$ of the baryon. The single-particle energy of the baryons, therefore, becomes a function of their momentum $k$ and effective masses: $E^*=\sqrt{k^2+m_b^{* 2}}$.

This approach describes parity doubling in the baryon octet implying a mass splitting between the baryon parity partners which is assumed to be generated by the scalar mesonic fields $\sigma$ and $\zeta$~\cite{Detar:1988kn,Zschiesche:2006zj,Aarts:2017rrl,Sasaki:2017glk}. As a consequence of this nontrivial coupling, the effective nucleon mass never drops significantly below its vacuum expectation value, for nuclear saturation density it reaches a value of $m^{*}_N(\rho=\rho_0)= 816 $ MeV.

The chiral field dynamics are determined self consistently by the scalar meson interaction potentials, driving the spontaneous breaking of the chiral symmetry:
\begin{eqnarray}
U_{\rm sc} & = & V_0 - \frac{1}{2} k_0 I_2 + k_1 I_2^2 - k_2 I_4 + k_6 I_6   \nonumber \\
& + & k_4 \ln{\frac{\sigma^2\zeta}{\sigma_0^2\zeta_0}} - U_{\rm sb} ,
\label{veff}
\end{eqnarray}
with
\begin{eqnarray}
    I_2 = (\sigma^2+\zeta^2)&,&~ I_4 = -(\sigma^4/2+\zeta^4),\nonumber\\
    I_6 &=& (\sigma^6 + 4\, \zeta^6)
\end{eqnarray} 
where $V_0$ is included to ensure that the pressure in vacuum  vanishes (i.e. $U_{sc}=0$ for $T=0$ and $\mu_B=0$). The terms $I_n$ correspond to the basic building blocks of possible chiral invariants that form different meson-meson interactions. The logarithmic term in equation (\ref{veff}) introduced in Refs.~\cite{Heide:1993yz,Papazoglou:1996hf}, contributes to the QCD trace anomaly and is motivated by the form of the QCD beta function at the one-loop level. In addition, an explicit symmetry-breaking term is introduced in the scalar potential:

\begin{equation}
U_{\rm sb} = m_\pi^2 f_\pi\sigma +\left(\sqrt{2}m_K^ 2f_K-\frac{1}{\sqrt{2}}m_\pi^ 2 f_\pi\right)\zeta\,.
\label{vsb}
\end{equation}

The mean-field vector repulsion is mediated by the fields: $\omega$ for repulsion at finite baryon densities, the $\rho$ for repulsion at finite isospin densities, and the $\phi$ for repulsion when finite strangeness density is present. The vector fields depend on the respective conserved charge densities and are controlled by the potential $U_{\rm vec}$,
\begin{eqnarray}
U_{\rm vec}&=& -\frac12\left(m_\omega^2\omega^2 + m_\rho^2\rho^2 + m_\phi^2\phi^2\right)\nonumber\\ &-&g_4\left(\omega^4+6\beta_2\omega^2 \rho^2+ \rho^4+                 \frac12\phi^4\left(\frac{Z_\phi}{Z_\omega}\right)^2 \right.\nonumber\\
&+&3\left.\left(
\rho^2+\omega^2\right)\left(\frac{Z_\phi}{Z_\omega}\right)\phi^2\right)\,.
\end{eqnarray}

Similar to the effective mass $m_b*$ which is modified by the scalar interactions, the vector interactions lead to a modification of the effective chemical potentials for the baryons and their parity partners:

\begin{equation}
    \mu^*_b=\mu_b-g_{\omega b} \omega-g_{\phi b} \phi-g_{\rho b} \rho
\end{equation}

Note that the coupling strengths of the nucleons and hyperons were chosen to reproduce nuclear binding energies as well as optical potentials in nuclear matter of $\approx -28$ MeV for the $\Lambda$ and $-18$ MeV for the $\Xi$ \footnote{Note that we did not impose constraints on the hyperon couplings from symmetry relations (see e.g. Ref.~\cite{Glendenning:2000jx}), which could be investigated in the future.}.

The remaining mesonic and noninteracting hadronic degrees of freedom are included in a form of Hadron Resonance Gas~(HRG) as a thermal heat bath according to their vacuum masses.

Altogether, the baryonic interactions allow for a reasonable description of nuclear matter properties. The coupling constants of the hadronic sector and the parameters of the effective potential for these fields (see Ref.~\cite{Steinheimer:2011ea} for details) are chosen such that the properties of nuclear matter are reproduced: ground state density $n_0\approx0.16\, \mathrm{fm}^{-3}$, binding energy per nucleon is $E_0/B\approx-15.2$~MeV, asymmetry energy $S_0\approx31.9$~MeV, and compressibility $K_0\approx267$~MeV. All fixed parameters and coupling constants used in the CMF model are summarized in Table~\ref{param-table} in the appendix.

The quark degrees of freedom are introduced as in the Polyakov-loop-extended Nambu Jona-Lasinio (PNJL) model~\cite{Fukushima:2003fw}, where their thermal contribution is directly coupled to the Polyakov Loop order parameter $\Phi$ \cite{Motornenko:2019arp}, the quark thermal contribution reads as:
\begin{eqnarray}
	\Omega_{\rm q}=&-&VT \sum_{q_{i}\in Q}\frac{d_{q_{i}}}{(2 \pi)^3}\int{d^3k} \frac{1}{N_c}\ln\left(1+3\Phi e^{-\left(E_{q_{i}}^*-\mu^*_{q_{i}}\right)/T}\right.\nonumber\\
	&+&\left.3\bar{\Phi}e^{-2\left(E_{q_{i}}^*-\mu^*_{q_{i}}\right)/T} +e^{-3\left(E_{q_{i}}^*-\mu^*_{q_{i}}\right)/T}\right)\,,
	\label{eq:q}
\end{eqnarray}
where the index ${q_{i}}$ runs through $u,d,s$ flavors. The antiquark contribution can be obtained by replacing $\mu^*_{q_{i}}\rightarrow-\mu^*_{q_{i}}$, and $\Phi\leftrightarrow\bar{\Phi}$. The Polyakov-loop order parameter $\Phi$ effectively describes the gluon contribution to the thermodynamic potential and is controlled by the temperature dependent potential~\cite{Motornenko:2019arp}:

\begin{eqnarray}
    U_{\rm Pol}(\Phi,\overline{\Phi},T) &=& -\frac12 a(T)\Phi\overline{\Phi} \\ \nonumber
	 + b(T)\ln \Bigl[1-6\Phi\overline{\Phi}\Bigr.
	 &+& \Bigl. 4(\Phi^3+\overline{\Phi}^{3})-3(\Phi\overline{\Phi})^2\Bigr] \,, \\
    a(T) &=& a_0 T^4+a_1 T_0 T^3+a_2 T_0^2 T^2,  \nonumber \\
    b(T) &=& b_3 T_0^4  \nonumber\,.
\end{eqnarray}

The dynamical quark masses $m_q^*$ of the light and strange quarks are also determined by the $\sigma$- and $\zeta$- fields, with the exception of a fixed mass term $m_{0q}$, which can be understood as the contribution of the gluon condensate to the quark mass:
\begin{eqnarray}
m_{u,d}^* & =-g_{u,d\sigma}\sigma+\delta m_{u,d} + m_{0u,d}\,,&\nonumber\\
m_{s}^* & =-g_{s\zeta}\zeta+ \delta m_s + m_{0q}\,.&
\end{eqnarray}

The full grand canonical potential of the CMF  model is expressed as follows:
\begin{eqnarray}
\Omega=\Omega_{\rm q} + \Omega_{\bar{\rm q}} + \Omega_{\rm h} + \Omega_{\rm \bar{h}}  - \left(U_{\rm sc} + U_{\rm vec} + U_{\rm Pol}\right)\,,
\end{eqnarray}
$\Omega_{\rm h}$ and $\Omega_{\rm \bar{h}}$ are the contributions from the hadrons which are the octet and the parity partners according to $\cal {L_B}$ and the rest of the hadron list is incorporated in a form of a hadron resonance gas. Since we work in mean-field approximation, the kinetic term of the mesons is not included in the Lagrangian. The mesonic degrees of freedom  are included explicitly in the grand canonical potential as single-particle states, i.e., as a sum over the corresponding thermal Bose-integrals for all mesonic species using their vacuum masses. $U_{\rm sc}$ is the mean-field interaction potential of the scalar mean fields $\sigma$ and $\zeta$, and  $U_{\rm vec}$ of the repulsive vector mean fields $\omega$, $\rho$, and $\phi$. $U_{\rm Pol}$ describes an effective gluon potential contribution as a part of the PNJL description.

The transition between the quark and hadronic degrees of freedom is controlled by two mechanisms \footnote{Note in an earlier version of the CMF model which does not include the chiral partners of the baryons, the deconfinement phase transition is moderated by an additional $\Phi$ term in the effective mass of the fermions~\cite{Dexheimer:2009hi}. There the point-like hadrons are suppressed by an explicitly-$\mu_B$-dependent term in the Polyakov Loop potential.}:
\begin{enumerate}
\item As the Polyakov loop order parameter becomes finite, free quarks can appear.
\item Hadrons are suppressed in the deconfined phase due to the excluded volume interactions.
\end{enumerate}
The suppression of hadrons at high energy densities is maintained by their excluded-volume hard-core interactions~\cite{Rischke:1991ke,Steinheimer:2011ea}.
Due to the assumption of finite size the hadrons are attributed an explicit volume term. This volume term then introduces an effective chemical potential $\mu^{\rm eff}_j$, which replaces the hadron chemical potential used to calculate the thermal contributions in $\Omega_h$:
\begin{equation}
    \mu^{\rm eff}_j=\mu^*_j - v_j\,P\,,
\end{equation}
for each hadronic particle specie $j$. Here, $P$ is the total pressure of the system and the $v_j$ are the EV parameters for the different particle species. Note that, in the previous works, the CMF model assumed only two different values of $v_j$, namely:
\begin{itemize}
    \item $v_j=1~\mathrm{fm}^3$  for baryons;
    \item  $v_j=1/8~\mathrm{fm}^3$ for mesons;
\end{itemize}
while quarks are always assumed point-like.
Note that the EV interaction scheme relates the excluded volume parameter  $v$ (particle's proper volume) to the hard-core radius $R$ as $v=\frac{16}{3}\pi R^3$ which for $v=1$~fm$^{3}$ yields $R\approx 0.39$ fm. In other words, each hadron excludes a volume with twice its radius. These values of hard-core radius are in agreement with the analysis of nucleon-nucleon scattering phase shift data~\cite{Wiringa:1994wb} and are important for the  thermodynamic consistency of the model.

As soon as quarks contribute to the pressure $P$, they reduce the hadronic density $\rho_i$ by lowering their chemical potential:
\begin{eqnarray}
\rho_i=\frac{\rho^{\rm id}_i (T, \mu^*_i - v_i\,P)}{1+\sum\limits_{j{\rm \epsilon HRG}} v_j \rho^{\rm id}_j (T, \mu^*_j - v_j\,P)},
\end{eqnarray}
where $i$ refers to all possible contributions from baryons, mesons as well as quarks.

In its default version, the CMF model predicts two first-order phase transitions for isospin-symmetric matter. The nuclear liquid-vapor phase transition mimics the transition from dilute gas of nucleons to the dense nuclear matter, this transition is located at $\mu_B\approx m_N$ with critical temperature $T_{\rm CP}\approx 17$ MeV.
At higher densities, the CMF model exhibits a first-order phase transition due to the chiral symmetry restoration among baryon parity partners~\cite{Steinheimer:2011ea,Motohiro:2015taa} with rather low critical temperature $T_{\rm CP}\approx 17$ MeV. 
The transition occurs due to the rapid drop in the chiral condensates $\sigma$ and $\zeta$ so the mass gap between parity partners is reduced.

The CMF model can be applied to study neutron stars without changing its parameters. In this case, electric charge neutrality and $\beta$-equilibrium are  imposed so the conditions of neutron star interior are fulfilled. To model the NS crust, which presumably consists of mostly neutron rich nuclei and clusters in equilibrium, an additional input is needed. That is done by matching the classical crust-EOS~\cite{Baym:1971pw} to the CMF-EOS at $n_B\approx0.05~{\rm fm^{-3}}$, such that below this density the matter is described by the crust EOS.

In the current version, the CMF model includes a plethora of different hadronic states, going beyond the baryonic octet and their parity partners, to include full hadron list. In principle that means the hundreds of different hadron species, with all of them having different masses and charges, which could also have different couplings to the mean fields. In this work we intend to extract some systematical properties about the hadronic interactions by a comparison to lattice data and show how then these properties will affect the models predictions at finite $\mu_B$. All other interaction parameters of the model were constrained before and only the EV are varied as free parameters. Thus, our results will also highlight the robustness of the CMF approach to the changes of the hadronic interaction properties.

\begin{figure*}[t]
\centering
\includegraphics[width=1.0\textwidth]{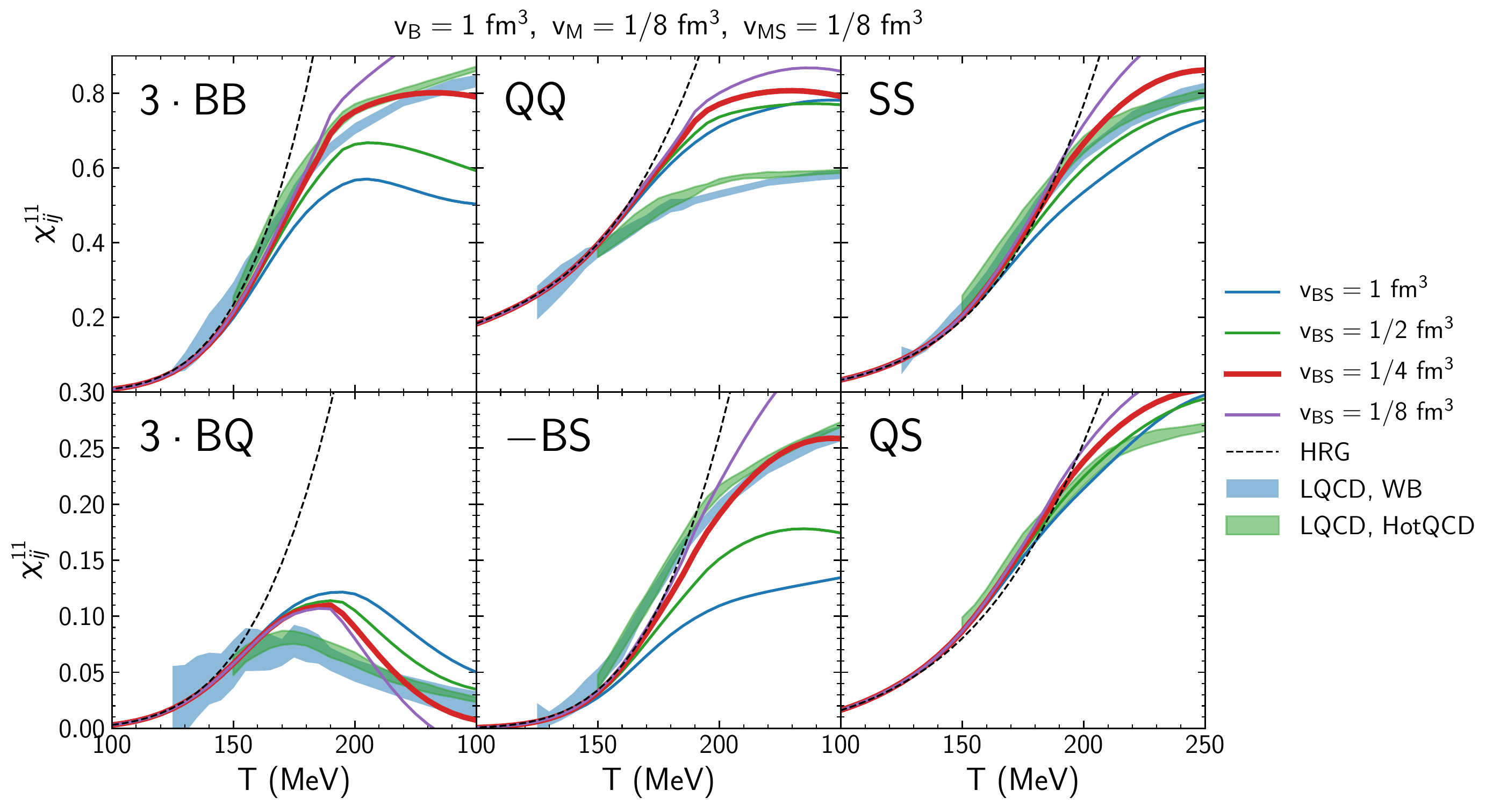}

\caption{Second-order susceptibilities $\chi^{11}_{ij}$ as functions of temperature $T$ for various excluded volume parameters of strange baryons $v_{\rm BS}$. The predictions of the CMF model, solid lines, are compared with available results of lattice QCD calculations by the Wuppertal-Budapest collaboration~\cite{Borsanyi:2011sw} and HotQCD collaborations~\cite{Bazavov:2012jq}, blue and green colorbands, respectively. The HRG results are also presented by the black dashed lines. All susceptibilities related to the baryon  number and strangeness show a strong sensitivity to the hyperon EV. The line which best fits to the lattice data is presented in bold for $v_{\rm BS}=1/4$~\fm.
}
\label{fig:susc}
\end{figure*}

\begin{figure*}[t]
\centering
\includegraphics[width=1.00\textwidth]{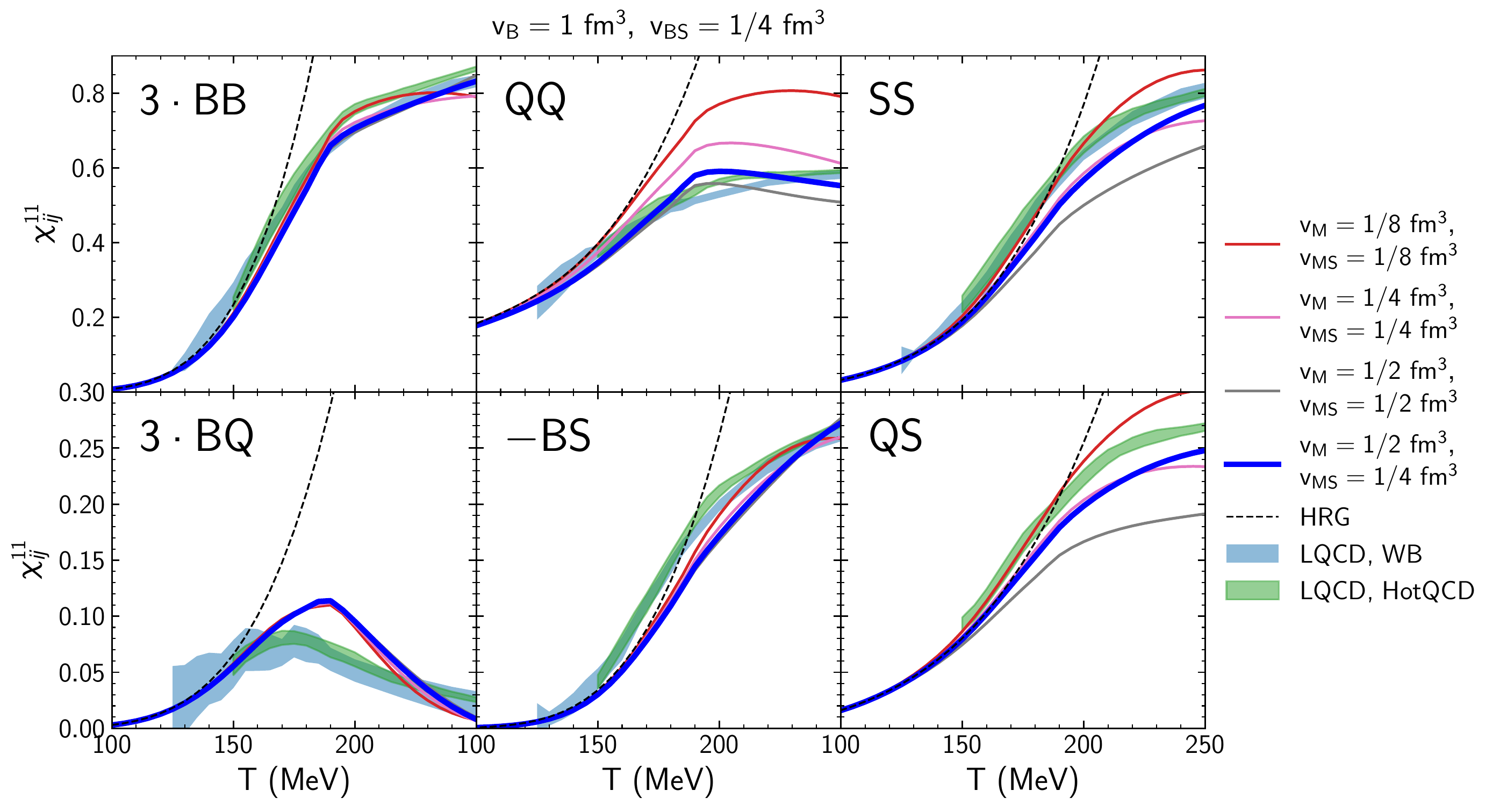}
\caption{The same as Fig.~\ref{fig:susc}, but for various excluded-volume parameters of mesons $v_{\rm M}$ and strange mesons $v_{\rm MS}$. The line which best fits to the lattice data is presented in bold for $v_{\rm M}=1/2$~\fm and $v_{\rm MS}=1/4$~\fm. The EV parameter of strange baryons is fixed to $v_{\rm BS}=1/4~{\rm fm^3}$ as the best result from Fig.~\ref{fig:susc}. The predictions of the CMF model, solid lines, are compared with available results of lattice QCD calculations by the Wuppertal-Budapest collaboration~\cite{Borsanyi:2011sw} and HotQCD collaborations~\cite{Bazavov:2012jq}, blue and green colorbands, respectively. The HRG results are also presented by the black dashed lines.  The electric charge susceptibilities show a particular strong dependence on the meson EV. Only the baryon-electric charge correlation appears to be insensitive.}
\label{fig:susc_mes}
\end{figure*}

\section{Lattice data comparison}
\label{sec:latice-analysis}

Important information about the phase structure at vanishing and finite baryon densities can be extracted from the fluctuations and correlations of conserved charges which are characterized by the susceptibilities~\cite{Koch:2008ia}, these quantities are sensitive to the effective degrees of freedom and their interactions. 
The critical regions of the QCD phase diagram are characterized by a nonmonotonic behavior of these susceptibilities~\cite{Stephanov:2008qz}.
However, the lattice results for vanishing chemical potentials, show a smooth transition between two baselines, a noninteracting ideal Hadron Resonance Gas (HRG) and weakly interacting quark-gluon matter in the region of temperatures between 100 and 250 MeV \cite{Borsanyi:2013bia,Bazavov:2014pvz}.
From duality arguments it should be possible to describe this transition, up to a certain point, in terms of a strongly interacting gas of hadronic degrees of freedom. Such a study is usually carried out by phenomenological models where the effective degrees of freedom and their interactions are given as input. 

Here we employ an effective hadron-quark model, the CMF model, which already incorporates a smooth transition between hadrons and quarks.
First, different second-order susceptibilities of conserved charges are calculated within the CMF model and a comparison with available lattice data is presented. This is done to highlight the importance of different repulsive interactions for non-strange and strange baryons and mesons on the extracted susceptibilities.
Thermal model analysis of experimental hadron yields already provides indications that flavor-dependent interactions in the EV-HRG are important to describe the transition region properly~\cite{Alba:2016hwx}. This idea was then further extended by a brief analysis of lattice QCD~(LQCD) susceptibilities in Ref.~\cite{Vovchenko:2017zpj}. The results indicated that susceptibilities which involve the baryon $B$ and strange $S$ charges are sensitive to the repulsive interactions among strange hadrons.
Using the CMF model we can study how the LQCD data can be described by a proper modeling of the repulsive short-range interactions represented by the effective excluded-volume sizes of the hadrons.

The conserved charge susceptibilities are related to the Taylor series expansion in powers of baryon, electric, and strange chemical potentials, $\mu_B$, $\mu_Q$, and $\mu_S$, of the thermodynamic pressure of matter at vanishing chemical potentials~\cite{Allton:2002zi}. The pressure expansion to finite chemical potentials takes the form
\begin{equation}
P = P_0 + T^4 \sum_{i,j,k} \frac{1}{i!j!k!} \chi^{i,j,k}_{B,Q,S} \left(\frac{\mu_B}{T}\right)^i
\left(\frac{\mu_Q}{T}\right)^j
\left(\frac{\mu_S}{T}\right)^k\,,
\label{eq:p_exp}
\end{equation}
where $P_0$ is the pressure at vanishing chemical potentials, and $\chi^{i,j,k}_{B,Q,S}$ are the conserved charge susceptibilities which are defined as:
\begin{equation}
\chi^{i,j,k}_{B,Q,S} = \frac{\partial^i \partial^j \partial^k P(T,\mu_B,\mu_Q,\mu_S)/T^4}{\partial \left({\mu_B/T}\right)^i \partial \left({\mu_Q/T}\right)^j \partial \left({\mu_S/T}\right)^k}\,.
\label{eq:chi}
\end{equation}
We limit the study only to second-order derivatives of the QCD pressure which already provides sufficient information to extract the hierarchy of EV sizes in the baryonic, strange, and mesonic sectors of hadronic matter.

Throughout all following results, we assume that the size of non-strange baryons is fixed to the size of the nucleon $v_{\rm B}=1$ \fm. This value is found to be in agreement with the microscopical quantum nuclear interactions of nucleons and is also supported by the analysis of LQCD data~\cite{Vovchenko:2017drx}. The value of $v_{\rm B}=1$ \fm corresponds to the proton radius as $R_{\rm p}=\sqrt[1/3]{\frac{3}{16\pi}v_{\rm B}}\approx0.39$~fm, the value is in agreement with the values suggested by the analysis of NN-scattering phase shift data~\cite{Wiringa:1994wb}. In the first step, the sensitivity of the susceptibilities on the strange baryon size is presented. The values of strange baryon sizes are varied as $v_{\rm BS}=1,1/2,1/4,1/8$ \fm. The volume of mesons here is initially fixed to $1/8$ \fm as in Ref.~\cite{Steinheimer:2011ea,Motornenko:2019arp} and will be varied later.

The resulting second-order susceptibilities are presented in Fig.~\ref{fig:susc}. As expected, the $BB$, $SS$, $BS$ susceptibilities show a strong sensitivity to the size of strange baryons in the temperature range $150<T<250$ MeV, which can be considered as the transition region between hadrons and quarks. A decrease of the strange baryon size, to $v_{\rm BS}=1/4$ \fm, allows a reasonable description of the $BB$, $SS$, and $BS$ susceptibilities.

The susceptibilities which involve the electric charge, however, show much less sensitivity to the strange baryon volume. Since a large fraction of the electric charge is carried by mesons, a change in the meson EV parameter should affect the electric charge susceptibilities. To study the susceptibilities which involve the electric charge, we vary the EV parameters for strange $v_{\rm MS}$ and non-strange $v_M$ mesons while $v_B =1$ \fm  and $v_{\rm BS}=1/4$ \fm are fixed as a result of the comparison presented in Fig.~\ref{fig:susc}. The results are presented in Fig.~\ref{fig:susc_mes} where four combinations of meson volumes are compared, $v_{\rm M}=v_{MS}=1/8$ \fm as in the default version of the CMF, $v_{\rm M}=v_{MS}=1/4$ \fm, $v_{\rm M}=v_{MS}=1/2$ \fm, and $v_{\rm M}=1/2$ \fm $v_{\rm MS}=1/4$ \fm. From these parametrizations the last one, which assumes a larger volume for non-strange mesons appears to describe the lattice data best.

Consequently, the parametrization with $v_B=1$ \fm, $v_{\rm BS}=1/4$ \fm, $v_{\rm M}=1/2$ \fm, $v_{\rm MS}=1/4$ \fm provides a much improved agreement with LQCD data for the second-order $BB$, $QQ$, $SS$, $BS$, $QS$, susceptibilities. Only the $BQ$ susceptibility appears to be unaffected by all EV parametrizations studied above. Since the $BQ$ combination is sensitive to the baryon charge correlations we conjecture that the $BQ$ susceptibility can be better described by a change of the EV parameters of the $\Delta$- and $N^*$ baryons. This would require one or more additional parameters related to the $\Delta$ and $N^*$ repulsive interactions, supporting the scenario of a unique EV parameter for every hadron, which are, however, mainly unknown. Such a picture seems reasonable and it would introduce a whole plethora of new parameters which allows for the description of even higher orders of LQCD susceptibilities.

\section{Consequences of the modified excluded volumes }
\label{sec:phase}

As discussed above, the introduction of species-dependent repulsive interactions of hadrons yields a good description of lattice QCD data, essentially up to an arbitrary order. Such a procedure, however, poses the question what conclusions can be drawn. Instead of trying to understand and justify every parameter, it is more convenient to study the sensitivity of the CMF model predictions for the high-density matter on these parameters. In the following, we will discuss how the modified EV parameters change the phase structure of the model and the equation of state for dense nuclear and neutron star matter. 
Thus, the goal of this section is to explain the consequences of the modified hyperon repulsion on different relevant states of matter: iso-spin symmetric, heavy-ion collisions (with strangeness conservation) as well as net strange matter. All these forms of QCD matter can be studied in different experimental and observational scenarios:

\begin{enumerate}
    \item \textit{Isospin Symmetric matter}: Here one assumes that up and down quarks (as well as protons and neutrons) are equally abundant. This scenario is often studied when one refers to the 'QCD-phase diagram'. In particular we assume that the strange chemical potential $\mu_S=0$ vanishes which can lead to a finite net strangeness.
    \item \textit{The EOS for Heavy Ion Collisions}: This state of matter is close to the isospin symmetric matter, but obeys an additional constraint of zero net strangeness. This type of matter is created in heavy-ion collisions at various beam energies where net strangeness is conserved.
    \item \textit{Neutron star matter}:  Neutron stars are cold compact stellar objects which are composed of QCD matter in $\beta$-equilibrium and in local charge neutrality. At low densities neutrons are much more abundant than protons and strangeness is not conserved, i.e., $\mu_S=0$. The densities in the NS interiors surpass several times nuclear saturation density.
    The description of such matter is essential for the calculations of neutron star properties and stands as a benchmark for QCD phenomenology for a region in the QCD phase diagram which is not accessible by LQCD methods or heavy-ion collisions.
\end{enumerate}

\subsection{Phase structure of Isospin Symmetric matter}

\begin{figure}[t]
\centering
\includegraphics[width=.48\textwidth]{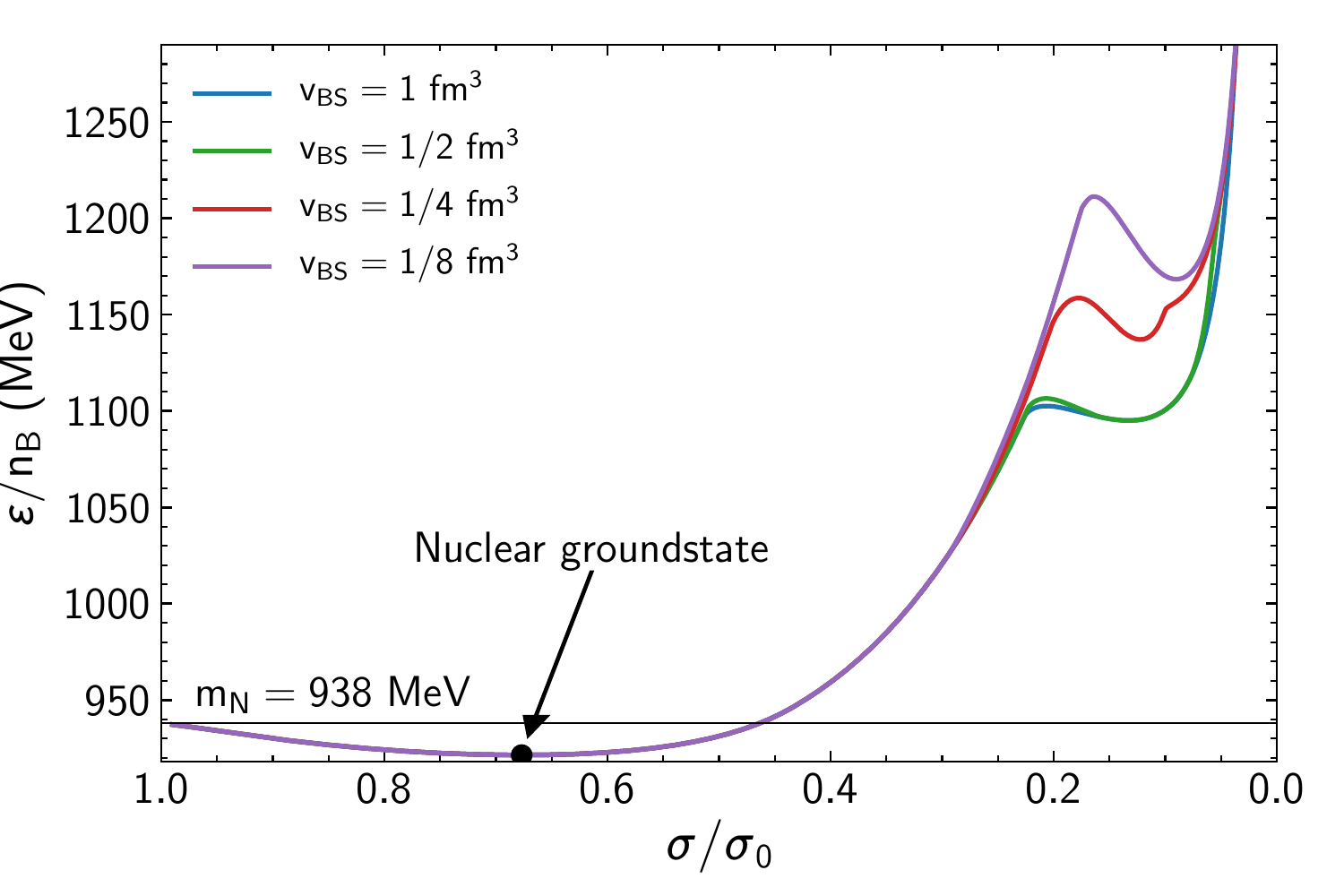}
\caption{Energy per baryon $\varepsilon/n_B$ as a function of the chiral condensate $\sigma/\sigma_0$ for $T=0$, isospin symmetric nuclear matter. Four different parametrizations of the strange baryon excluded volume $v_{\rm BS}=1,\,1/2,\,1/4,\,1/8~{\rm fm^3}$ are presented. While the nuclear ground-state properties are unchanged, a second minimum in the energy per baryon located at smaller values of the chiral condensate, indicating the chiral phase transition, is sensitive to the EV parametrization. This second minimum signals a metastable state of chirally restored matter.}
\label{fig:E_A_sigma}
\end{figure}

The interactions in the CMF model provide a reasonable description of the nuclear ground-state properties, such as binding energy, compressibility, asymmetry energy, and the slope parameter~\cite{Motornenko:2019arp}. The changes of the EV parameters introduced in the previous section do not allow hyperons to appear at below and slightly higher than the nuclear saturation density. As shown in Fig.~\ref{fig:E_A_sigma} the properties of the nuclear ground state are not affected by the change of EV parameters. The figure shows the energy per baryon $\varepsilon/n_B$ at $T=0$ for isospin symmetric matter as function of the order parameter, the chiral condensate $\sigma/\sigma_0$. After a density of  $n_B\approx 0.5$ \fmm the parametrizations start to deviate for different values of the hyperon EV parameter (at $T=0$ mesons are not excited and the meson Bose condensation is not included in our calculations). The nonmonotonic behaviour of the energy per baryon indicates the presence of a phase transition with a metastable state. Note that even though a metastable state with a small energy barrier is created, no absolutely stable state of matter can be generated by the appearance of the hyperons.

\begin{figure}[t]
\centering
\includegraphics[width=.48\textwidth]{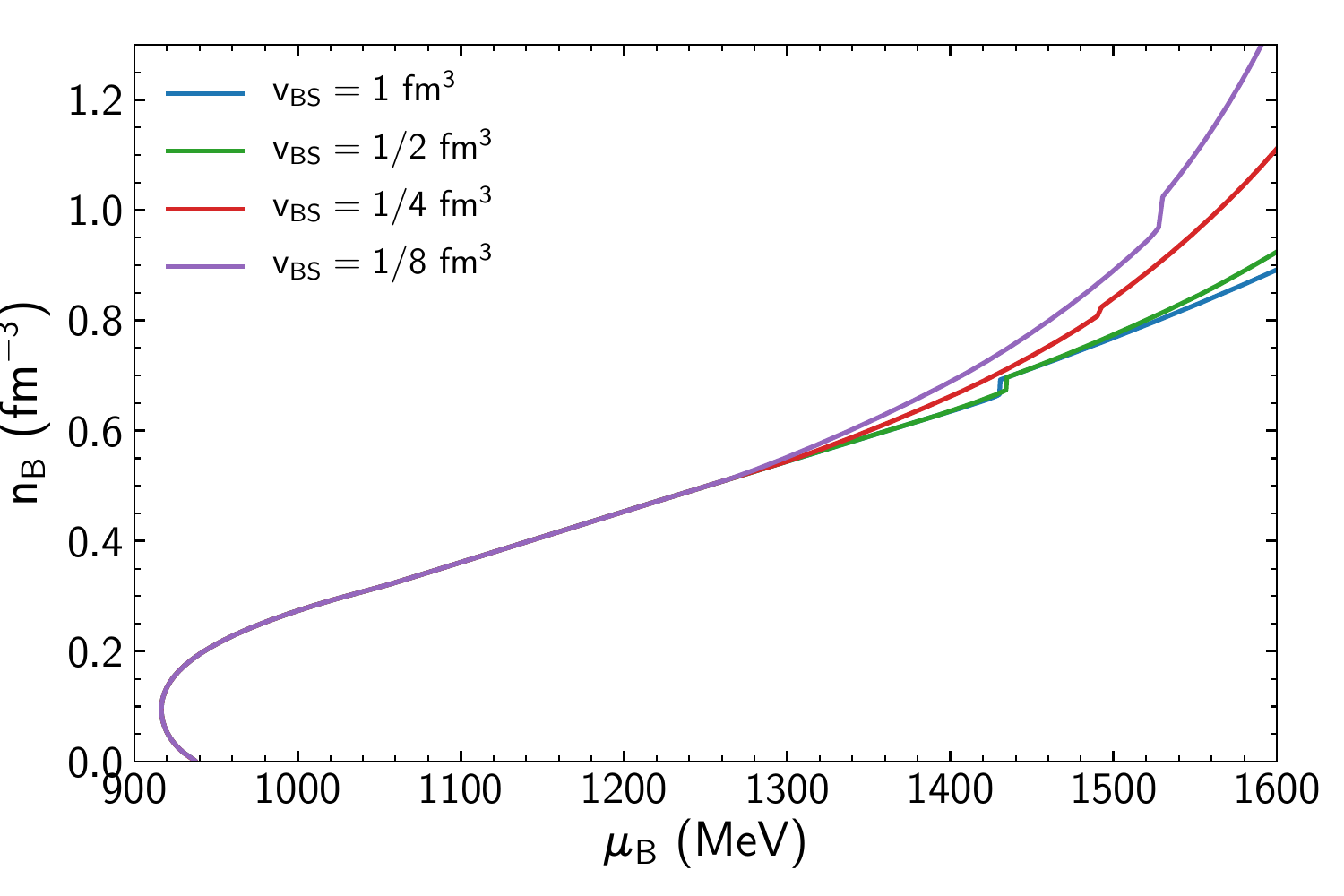}
\caption{Baryon density $n_B$ as a function of baryon chemical potential $\mu_B$ for $T=0$ isospin symmetric nuclear matter for four different parametrizations of strange baryon excluded volume $v_{\rm BS}=1,\,1/2,\,1/4,\,1/8~{\rm fm^3}$. Note that no additional phase transition appears while the chiral transition is shifted to higher values of baryon chemical potential.}
\label{fig:nB_muB}
\end{figure}

The effect of the modified EV parameters on the phase structure is also depicted in figure \ref{fig:nB_muB}. Here we show the net baryon density $n_B$ as a function of the baryon chemical potential $\mu_B$ for $T=0$ and isospin symmetric matter. To better illustrate the position of the first-order transition, a Maxwell construction between two coexisting phases was done. For the default version of the CMF a very weak chiral phase transition appears at $\mu_B^C \approx 1400$ MeV, with a critical endpoint at $T_{\rm CP}\approx 17$ MeV. As the volume of the hyperons is decreased, this transition gets slightly stronger, i.e., the latent heat and the jump in the density is increased and at the same time the critical chemical potential is increased. It was checked that the value of critical temperature (temperature of the chiral critical point) is not significantly affected by the change of interaction parameters. Thus, the general characteristics of the phase structure, i.e., a critical endpoint at a very low temperature, are not changed. 

The standard CMF parametrization, Refs.~\cite{Steinheimer:2010ib,Steinheimer:2011ea,Motornenko:2019arp}, yields matter at $T=0$ (assuming $\mu_S=0$), which is only composed of nucleons and their parity partners. Heavier hadrons as deltas and hyperons are suppressed by the interactions.
This is a result of the EV interactions in the CMF model: hadrons are suppressed at higher densities as a result of their repulsive hard-core interactions. The quarks become the dominant degrees of freedom in the medium. The degree of suppression depends on the repulsion coefficient, i.e. the EV parameter. The higher the value of this parameter, fewer hadrons will be present as the pressure is increased. If the EV coefficient of the strange baryons is smaller than the EV of the non-strange baryons then the strange baryons will survive to higher energy densities.

This allows for a distinct type of nuclear matter to emerge prior to the transition to the quark matter. Hyperonic matter thus appears as an additional phase between nuclear and quark matter. Hypermatter is a metastable state which appears as an exotic strange form of matter~\cite{Schaffner:1992sn,Schaffner:1993nn,Gilson:1993zs,SchaffnerBielich:1996eh,Scherer:2008zz,Steinheimer:2008hr,Botvina:2014lga}.

\begin{figure}[t]
\centering
\includegraphics[width=.48\textwidth]{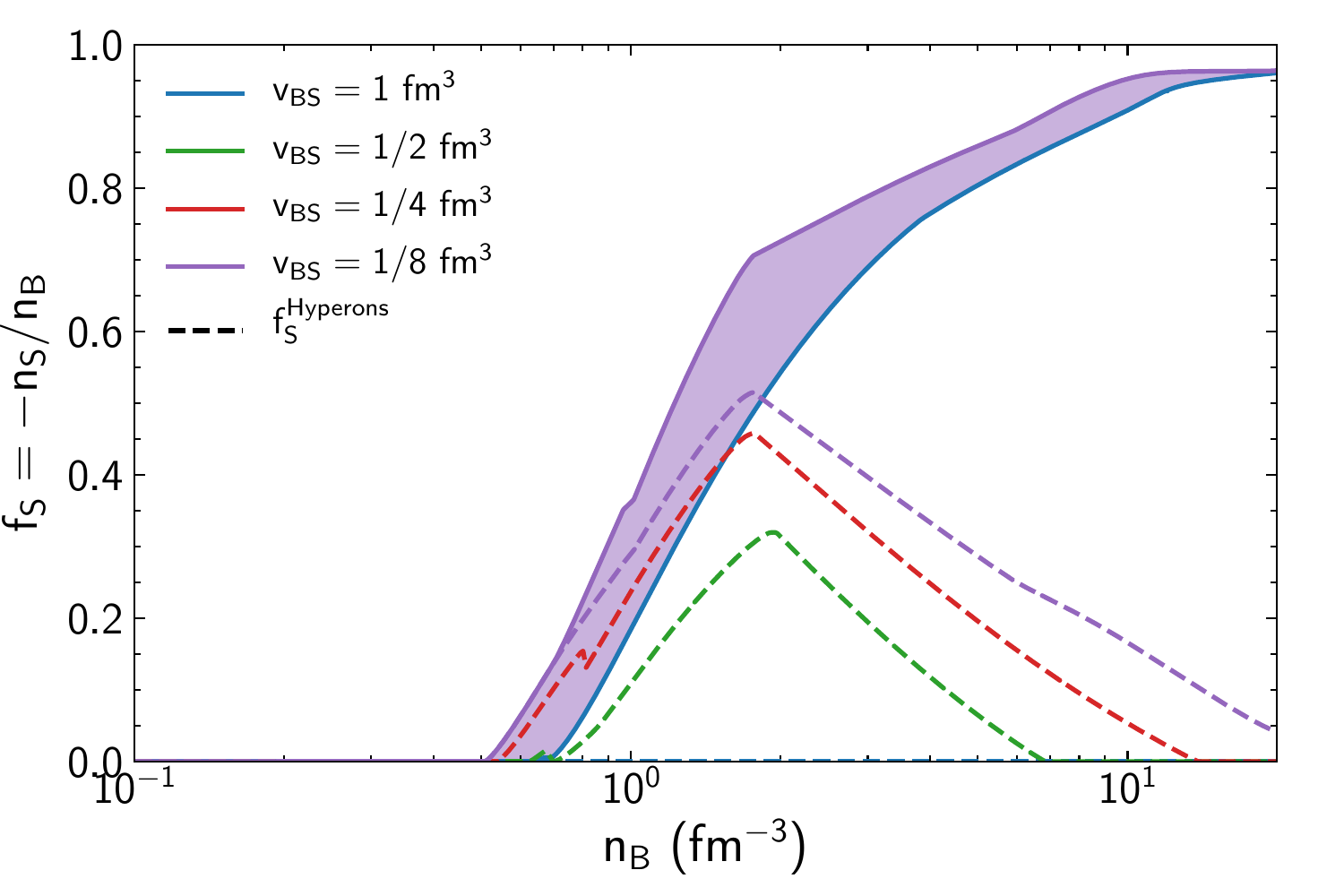}
\caption{Strangeness fraction $f_S$ as a function of baryon density $n_B$ for $T=0$ isospin symmetric nuclear matter. The band illustrates the range of values due to varying the hyperon EV parameter in range $v_{\rm BS}=1/8\textup{--}1$~fm$^3$. The hyperon contribution to strangeness fraction is illustrated by dashed lines. Note, no additional phase transition or a state of bound hyperon matter appear.}
\label{fig:fs_nB}
\end{figure}

Figure~\ref{fig:fs_nB} shows the strangeness per baryon $f_S=-(n_S/n_B)$ as a function of the baryon density for $T=0$ isospin-symmetric matter. The limit for $f_S$ is $3$ as then the matter would be made up completely of strange quarks. A value of $f_S=1$ would correspond to $\Lambda$ matter where $1/3$ of the baryon charge is carried by the strange quarks.

The four values of the hyperonic volume are located within the purple band which covers the possible range of $f_S$ for $v_{\rm BS}=1/8\textup{--}1$ \fm. The dashed lines correspond to the fraction of $f_S$ which stems from the hyperons. Since in the scenario with $v_{BS}=1$ fm$^3$ all strangeness is carried by the $s$-quark, the blue dashed line constantly stays at zero. As the EV of the strange hadrons is decreased, the fraction of hyperonic matter is increased significantly.
At the density around $n_B\approx1.5\textup{--}2$ \fmm The hyperons start to be suppressed, this is a result of EV suppression when the free quarks create a significant contribution to the total system pressure.
At very high densities, $n_B\approx 20$ \fmm, strangeness fraction $f_S$ for all parametrizations coincide, this is where the pure quark matter is produced and all hadrons are completely suppressed. However, a super-rich strange state $f_S>1$ is never produced by multistrange baryons and the strangeness fraction increases continuously from 0 to 1, which is also the limit for a free gas for three quark flavors.

As the metastable states observed in figure \ref{fig:E_A_sigma} appear for systems below the critical temperature of $T_{CP}\approx 17$ MeV, states of hyperon rich matter may survive here for an extended time.
It is questionable whether such a cold and dense environment could be created in heavy-ion collisions; however, in neutron stars and their mergers this scenario appears feasible.

\subsection{The EOS for Heavy Ion collisions}

A direct confirmation of the equation of state at high density from heavy-ion data would require the space-time simulation of the collision, using the CMF model as input. A first study in this direction was already done for low beam energies using ideal fluid dynamics~\cite{Seck:2020qbx}. However, for heavy-ion collisions it is essential to also take into account the nonequilibrium aspects. Early studies extracted an effective nuclear equation of state from the flow data~\cite{Danielewicz:2002pu}. This method can not be directly compared with the CMF finite temperature EoS. It is planned to apply the proper treatment to take into account the interactions in relativistic transport through the mean-field description at finite temperatures, as shown in Refs.~\cite{Nara:2019qfd,Nara:2020ztb}.

This work focuses on thermodynamic properties of the CMF model related to heavy-ion collisions. Effects of the different EV parametrizations may be observed  in the late stages of heavy-ion collisions and neutron stars. The change in the repulsive properties leads to different thermodynamic properties of the system at the chemical freeze-out which,  potentially, can be measured through the final particle yields. The chemical freeze-out conditions depend on the energy of the nuclear collision, which allows experiments to probe various regions of the QCD phase diagram experimentally. Since the bulk evolution, at any given beam energy, is well characterized by the produced entropy per baryon, the mapping between the collision energy and the expansion path through the phase diagram can be done by the chemical composition of hadrons after the chemical freeze-out~\cite{Cleymans:1992zc,BraunMunzinger:1996mq,Becattini:2000jw}. For simplicity, we use the so-called freeze-out line for our comparison. Through the measured chemical composition of particles this line provides a mapping of the collision energy $\sqrt{s_{NN}}$ with temperature $T$ and baryon chemical potential $\mu_B$ at the chemical freeze-out. Here the chemical freeze-out curve from Ref.~\cite{Vovchenko:2015idt} is used.

\begin{figure}[t]
\centering
\includegraphics[width=.48\textwidth]{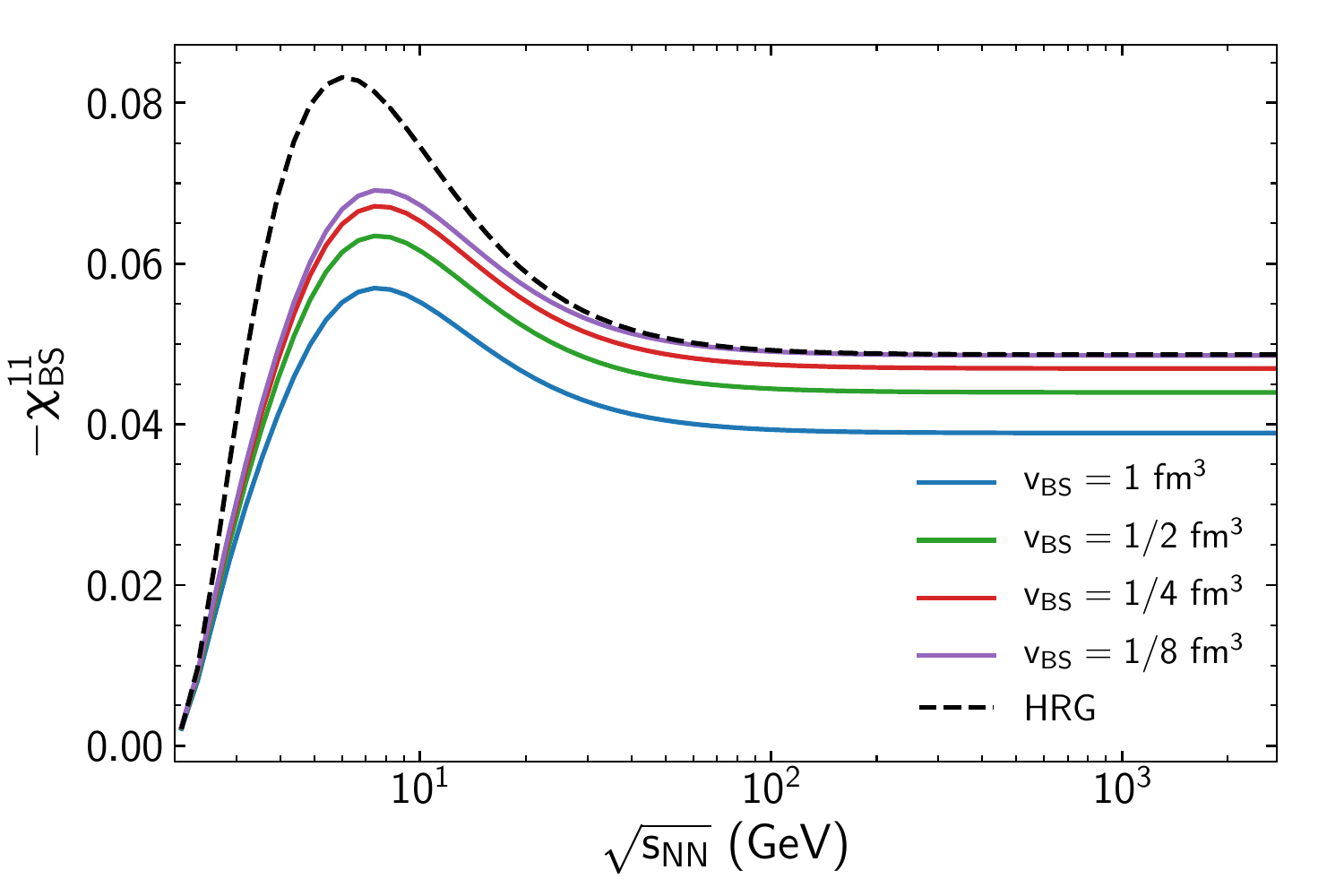}
\caption{Collision energy dependence of baryon-strangeness second-order susceptibility $\chi^{11}_{\rm BS}$ estimated in the CMF model along the chemical freeze-out curve of Ref.~\cite{Vovchenko:2015idt} for zero net-strangeness isospin-symmetric matter.}
\label{fig:BS_snn}
\end{figure}

The strangeness-baryon cross susceptibility $\chi^{11}_{BS}$ is particularly sensitive to the strange hadron EV-parameter: we estimate the values of $\chi^{11}_{BS}$ along the freeze-out line for four different values of $v_{\rm BS}$. Figure~\ref{fig:BS_snn} compares these resulting CMF-susceptibilities $\chi^{11}_{BS}$ with the well-known ideal HRG results. As the chemical freeze-out is assumed to occur when matter is quite dilute, moderate effects of the EV interactions are observed.  \footnote{ A more elaborate scenario for chemical freeze-out which implies two or more separate freeze-out points for strange and non-strange particles finds that strange hadrons could freeze-out at 10 to 15 MeV higher temperatures than the light hadrons at the highest collision energies~\cite{Steinheimer:2012rd,Bellwied:2018tkc,Flor:2020fdw}. A strange freeze-out at these higher temperatures could provide stronger signals of different EV interaction schemes.}

The matter at the studied freeze-out scenario does not produce such significant sensitivity to the EV parameters as the lattice QCD data. In addition, for a meaningful comparison of measured susceptibilities with our model calculation, some elaborate simulations, taking into account effects of the finite size and lifetime of the system, would be necessary. Thus we conclude that low baryon densities offer for the LQCD data a good benchmark to probe hadronic interactions.
These interactions and the related phase structure should be tested with heavy-ion collisions in the high baryon density regime, e.g., at FAIR facility. In addition, nuclear astrophysics offers an alternative venue through the study of neutron star properties and binary neutron star mergers with their gravitational wave signals~\cite{Most:2018eaw,Weih:2019xvw,Hanauske:2019qgs}.

\begin{figure*}[t]
\centering
\includegraphics[width=.95\textwidth]{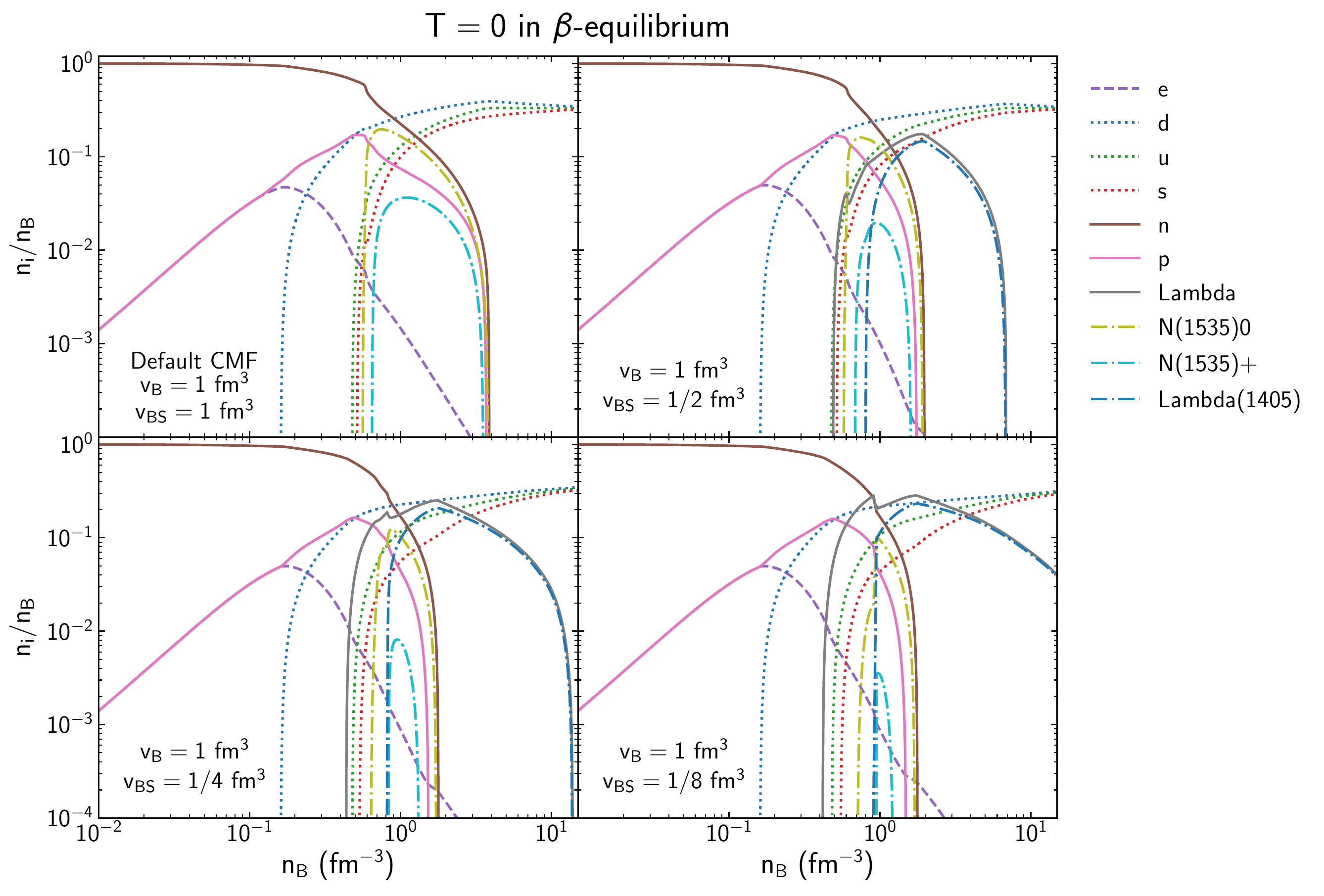}

\caption{Particle composition by the CMF model for $T=0$ matter in $\beta$-equilibrium. Particle number densities $n_i$, over the baryon density $n_B$ are presented as functions of baryon density, note that an additional factor $1/3$ for quarks is used. Four plots correspond to four different parametrizations of strange baryon excluded volume $v_{\rm BS}=1,\,1/2,\,1/4,\,1/8~{\rm fm^3}$. Electrons and baryons only from the groundstate octet are presented by solid lines, quarks by dotted lines, octet parity partners by dot-dashed lines.}
\label{fig:content}
\end{figure*}

\subsection{Neutron stars}
\label{sec:NS}

Observations of neutron stars provide another way to probe the equation of state of cold and dense nuclear matter and possibly deconfined quark matter. The CMF model, in its default parametrization, gives a satisfactory description of the properties of cold static nonrotating neutron stars. In particular, the mass-radius relation $M(R)$, and the tidal deformability $\Lambda$~\cite{Motornenko:2019arp}. 
The mass-radius relation is obtained by solving the Tolmann-Oppenheimer-Volkoff (TOV) equation~\cite{Tolman:1939jz,Oppenheimer:1939ne}, which uses the equation of state as input and provides density and pressure profiles of the NS. A solution of the TOV equation relates the central density to the NS mass and radius. The densities in the NS's interiors can reach several times nuclear saturation density $n_0$. This allows for the formation of quark cores in the interior of the stars~\cite{Alford:2015gna,Zacchi:2016tjw,Montana:2018bkb,Most:2018eaw,McLerran:2018hbz,Annala:2019puf,Jakobus:2020nxw,Tan:2020ics}. These cases are not yet observed. They could be tested in future by measurements of NS masses and radii, e.g., with the NICER X-ray telescope~\cite{Riley:2019yda,Miller:2019cac}, and by the next generation GW detectors~\cite{Punturo:2010zza,Evans:2016mbw}. However, even in the hadronic part of the EOS the chemical composition is not well known. The assumption of $\beta$-equilibrium implies that the matter is dominated by neutrons at densities close to $n_0$ and that the charge of the small admixture of protons is compensated by the same number of electrons~\footnote{Note that we have checked that the inclusion of muons does not alter our results in any significant way.}. With increasing density, heavier hadrons should appear. The implications of the hyperon appearance for the NS properties is actively discussed as the "hyperon puzzle", (for a review we refer to Ref.~\cite{Vidana:2018bdi}), which traces back to the 1960s~\cite{Saakyan:1960}.

Furthermore, in neutron star matter, $d$-quarks are favorable as compared with $u$-quarks or protons due to their opposite electric charge. They are easier to excite than neutrons. Even at $n_B\approx 2\,n_0$ the free quarks make up only to 20\% of the baryonic charge. In the CMF model, the chemical composition for isospin symmetric matter contains no free quarks for $n_B<2\,n_0$, and free quarks make up 20\% of the total baryon number only at $n_B \approx 5\, n_0$.

\begin{figure}[t]
\centering
\includegraphics[width=.48\textwidth]{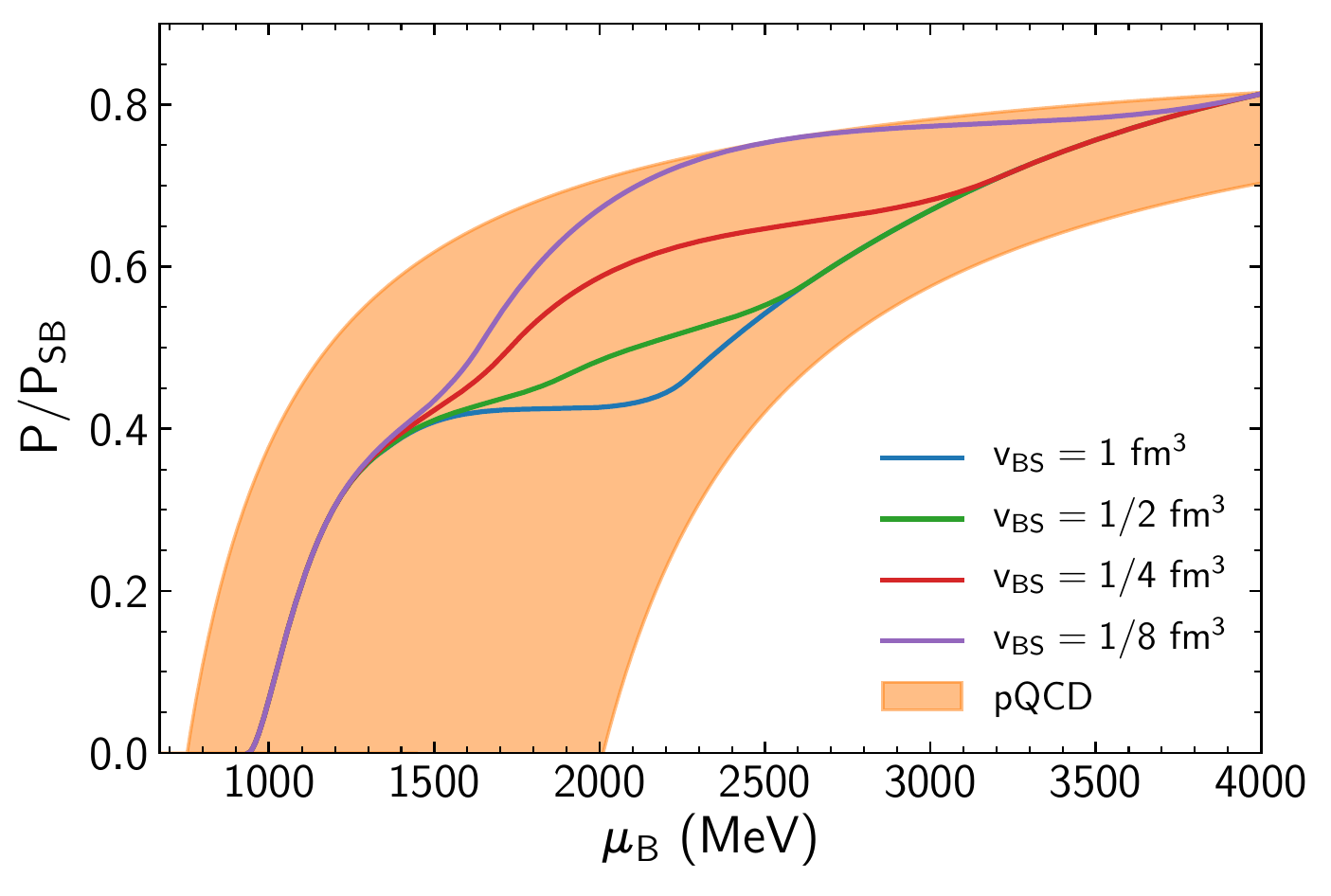}

\caption{The ratio of charge neutral CMF matter pressure $P$ at $T=0$ in $\beta$-equilibrium, $P$, to the Stefan-Boltzmann pressure limit of a massless 3-flavor gas of quarks, $P_{\rm SB}$.  Lines correspond to the four different excluded volume parameters of strange baryons used. The respective particle contents are illustrated in~Fig.~\ref{fig:content}. 
The yellow colorband illustrates parametrization~\cite{Fraga:2013qra} of three-loop pQCD calculations for pressure of cold quark matter in $\beta$-equilibrium~\cite{Kurkela:2009gj}.
}
\label{fig:pQCD}
\end{figure}

When the hyperon volumes are treated as the same as the non-strange baryons, i.e., $v_{\rm SB}=v_{\rm B}$, the hyperons in the CMF model are suppressed by both, their higher masses and their EV interactions. Hence, they do not appear in neutron stars~\cite{Motornenko:2019arp}. The same is true for any other higher mass baryons which are suppressed at $T=0$ by their repulsive excluded volume interactions. In this case, the hadronic part of the NS is only composed of nucleons and their parity partners.
The early appearance of the parity partners, as opposed to, e.g. the Delta baryons which have a smaller vacuum mass, is the parity doubling due to chiral symmetry restoration. Both the $N$ and $\Lambda$ as well as their parity partners will have a smaller effective mass than slightly heavier hyperons, or $\Delta$ ground-state baryons.

This scenario can significantly change as the EV-parameter of the hyperons is reduced: the chemical composition of neutron star matter is shown in Fig.~\ref{fig:content} as a function of the baryon density. The non-strange baryon EV parameter is fixed at $v_{\rm B}=1$ \fm, while the EV parameter of strange baryons is varied, $v_{\rm BS}=1,\,1/2,\,1/4,\,1/8$ \fm. 
Decreasing the hyperon repulsion allows $\Lambda$-baryons and their parity partners $\Lambda(1405)$ to populate the NS matter, while heavier hyperons are suppressed due to their higher mass. The threshold for the appearance of the $\Lambda$ corresponds to $n_B\approx 0.3-0.4$ \fmm (for all values of $v_{\rm BS}$ except the largest). This is clearly below the density of the chiral phase transition in the CMF model. The location of the chiral transition is sensitive to $v_{\rm BS}$ as well: for $v_{\rm BS}=1$ \fm and $1/2$ \fm the transition is located at $n_B\approx0.6$ \fmm. For $v_{\rm BS}=1/4$ \fm it is shifted to higher density, $n_B\approx 0.8$ \fmm. The transition is located at $n_B\approx1$ \fmm for $v_{\rm BS}=1/8$ \fm. At the chiral transition, the parity partner mass drops to the $\Lambda$-mass. Hence, $\Lambda(1405)$ contributes to the strangeness fraction similarly to the octet $\Lambda$-hyperon.

The reduced $v_{\rm BS}$-repulsion in the strange baryon sector yields a significant hyperon fraction of the total baryon density.
If the repulsion among the strange particles is eight times smaller than among non-strange, as illustrated in Fig.~\ref{fig:content} which shows that for $v_{\rm BS}=1/8$ \fm, the hyperons can survive up to extreme densities of 10 \fmm and even more. At these densities quarks are the dominant degrees of freedom.  However this type of matter is distinct from the quark matter due to the small admixture of strange hadrons. 

The appearance of the additional hyperon degrees of freedom leads to a softening of the NS-matter EOS. This inevitably changes the properties of neutron stars. To illustrate the change of the EOS due to the $v_{\rm BS}$, Fig.~\ref{fig:pQCD} shows the pressure $P$ for the CMF calculations as a function of the baryon chemical potential $\mu_B$ as compared with the results of pQCD calculations~\cite{Kurkela:2009gj}. The additional degrees of freedom at a given chemical potential yield additional pressure. For the values $v_{\rm BS}=1/4,\,1/8$ \fm a significant increase in $P/P_{\rm SB}$ is observed. This is a result of the sudden appearance and subsequent suppression of hyperons in the EOS. For $v_{\rm BS}=1/8$ \fm the increase reaches the borders of the pQCD bands of confidence suggesting that it can be considered as an absolute lower bound for $v_{\rm BS}$. However, all the parametrizations fit within the pQCD band and merge into one line at the region of chemical potential $\mu_B>3500$ MeV where the pQCD bands become narrow, there the baryon densities are extreme with $n_B>20$ \fm. At these values of $\mu_B$, as predicted by the CMF model, the matter is composed of free quarks only without admixture of hadrons. At the lower values of the chemical potential the pQCD bands permit various scenarios of hadron-quark interactions as shown by the CMF results.

\begin{figure}[t]
\centering
\includegraphics[width=.48\textwidth]{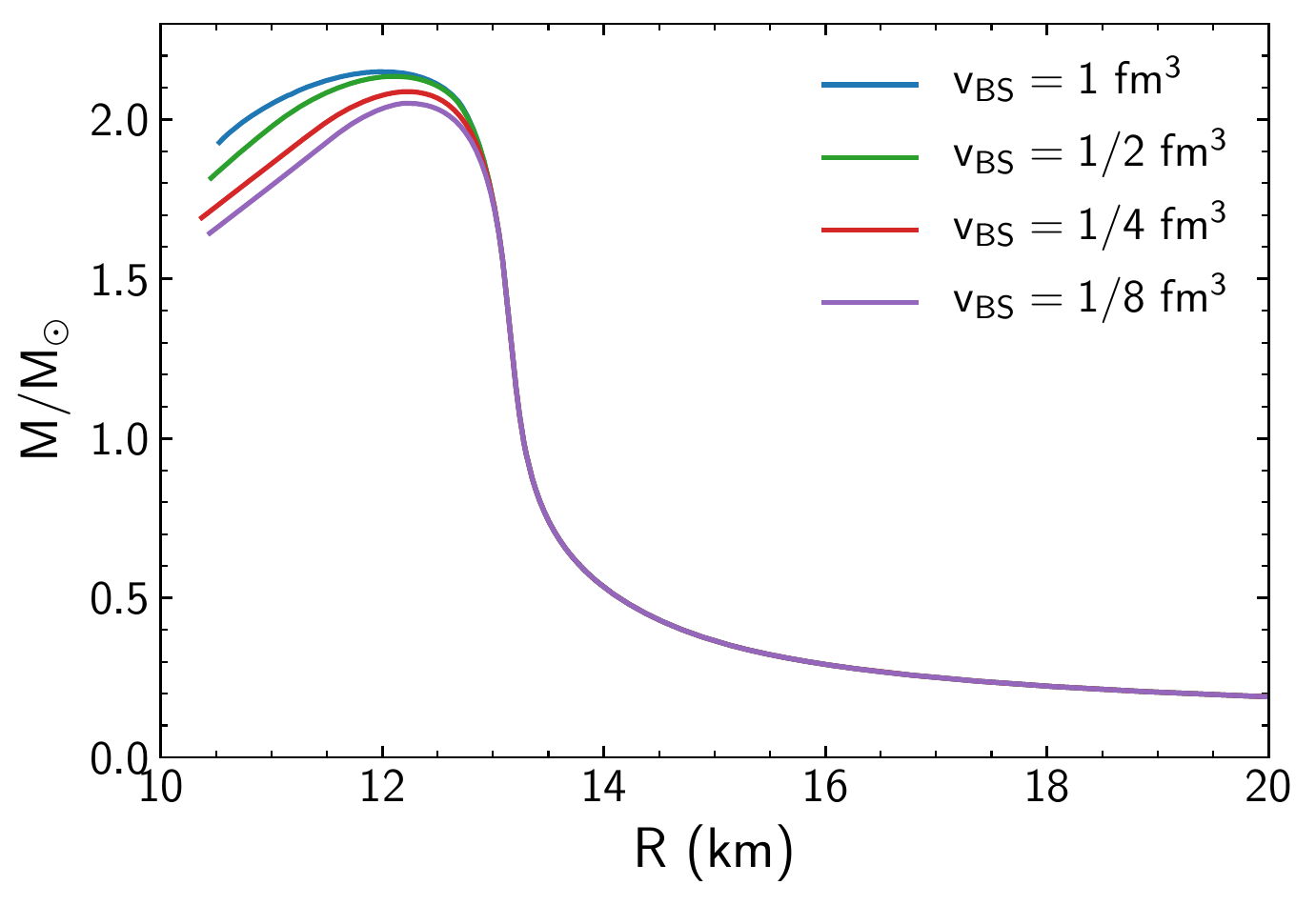}

\caption{The CMF mass-radius diagrams for four different excluded volume parameters of strange baryons. The respective particle content is illustrated in~Fig.~\ref{fig:content}. Note that the appearance of numerous strange baryons only slightly changes the mass-radius diagram, substantially affecting only the unstable branch of the solutions.
}
\label{fig:mr}
\end{figure}

These differences of the EOS due to the possible variation of $v_{BS}$ change the properties of neutron stars:  Fig.~\ref{fig:mr} depicts the mass-radius relations as calculated from the Tollman-Oppenheimer-Volkov equation\footnote{The numerical solutions were obtained using the TOV solver of~\cite{tovsolver}.}, for the EOS parametrizations discussed above. The additional degrees of freedom results in a softening of the EOS. The maximum central densities reached in the CMF model lie between 5 and 6 times nuclear saturation density. The softening decreases the maximum mass of the NS families by 5\%, from $M^{\rm max}\approx 2.15 M_\odot$ to $M^{\rm max}\approx 2.05 M_\odot$. The differences appear only in the highest mass region where the hyperons can influence the EOS. These high densities can be reached only in the most massive stars. 
For all parametrizations, the calculated properties of neutron stars, like the mass-radius relation, the chemical composition of the stars, and the tidal deformabilities are in good agreement with recent experimental constraints~\cite{Nathanail:2021tay,Most:2018hfd,Abbott:2018wiz,Hinderer:2009ca}. 
Since only mergers of neutron stars yield much higher densities and high temperatures, a study of the effects of the hyperonic repulsion in simulations of neutron star mergers is needed. Another worthwhile study could be the effect on neutron star cooling: an early study of the CMF model in the context of parity doubling showed that the cooling curve can be reasonably well described within this model. There, certain assumptions on the role of the parity partners are made~\cite{Dexheimer:2012eu}.

\section{Summary and Conclusions}
\label{sec:summary}

The CMF model is employed to study the change of hadronic properties and the EOS dependence on repulsive interactions. A decrease of the excluded volume parameters of hyperons and both strange and non-strange mesons, as compared with non-strange baryons, leads to an improved description of lattice QCD data of the second-order susceptibilities $\chi^{11}_{ij}$.

The structure of the conserved charge susceptibilities in the transition region of QCD is reflected in the complex structure of hadronic interactions. As the hadronic interactions appear to drive the susceptibilities towards the Stefan-Boltzmann values of a free gas of quarks, it is not possible to define a sharp transition point at which quark degrees of freedom 'appear' and hadrons 'disappear' using the susceptibilities alone.
    
Our improved comparison to the lattice QCD data shows that the hard-core repulsive interactions of hyperons is systematically smaller when using a smaller hard-core radius than for non-strange baryons. Also non-strange and strange mesons appear to have a smaller effective size than baryons. This provides the following scenario of the hierarchy of the repulsive hadronic interactions: non-strange baryons are more repulsive than non-strange mesons, while strange baryons and strange mesons are the least repulsive. These results can be further improved by attributing a separate repulsive parameter for each hadron specie. Since recent lattice QCD calculations also show that the $\Delta$ and $\Omega$ baryons show a parity doubling behavior, it will be worthwhile to extend our approach in the future to include also the higher mass states of the baryon decuplet.
    
Even though the interaction parameters between the hadrons are changed to better match the lattice QCD results, the phase structure at large baryon densities remains remarkably unchanged. The critical endpoint of the chiral transition therefore appears out of reach of heavy-ion experiments.
    
The properties of neutron stars are only weakly influenced by the hyperonic interactions, as they do get strong at the highest densities, as can only occur in relativistic collisions of NS. Consequently, the properties of static neutron stars and the properties of the QCD phase transition in cold NS matter seem to be dominated by the non-strange baryon degrees of freedom. The existence and location of the QCD phase transition is, therefore, constrained most strongly by neutron star observations and their binary mergers, as well as by nuclear matter properties, rather than by lattice QCD results.

The present analysis suggests that non-strange baryons have an EV parameter of $v_B=1$ \fm, the non-strange mesons volume is close to $v_M=1/2$ \fm. Strange baryons and strange mesons seem to have smaller excluded volume and therefore are less repulsive with an EV parameter of $v_{BS}=v_{MS}=1/4$ \fm. 
As the excluded volume mechanism in the CMF model provides the suppression of hadrons at high energy densities, the reduction of the strange baryon EV parameter means that hyperons will be less suppressed at higher energy densities than non-strange hadrons.

Our study implies that the most stringent constraints for the high density QCD equation of state can only come from binary neutron star mergers or heavy-ion collisions. The baryon densities reached exceed those found in cold neutron stars and the finite temperatures allow the excitation of the mesonic as well as the hyperonic degrees of freedom. The CMF model provides a unique framework to be used in simulations of neutron star mergers as well as heavy-ion collisions, thus providing an important link between high baryon density physical processes and lattice QCD data.\\

\begin{acknowledgments}

We thank Roman Poberezhnyuk for carefully reading the manuscript and fruitful discussions.
The authors are thankful for the support from HIC for FAIR and HGS-HIRe for FAIR.
The authors appreciate support by the Alexander von Humboldt (AvH) foundation and the BMBF through a Research Group Linkage programme. SP thanks University Grants Commission (India) for support. JS thanks the Samson AG and the BMBF through the ErUMData project for funding. 
JS and HSt thank the Walter Greiner Gesellschaft zur F\"orderung der physikalischen Grundlagenforschung e.V. for support. 
HSt acknowledges the support through the Judah M. Eisenberg Laureatus Chair at Goethe University.
\end{acknowledgments}

\appendix*
\section{Table of parameters}

\begin{table}[h!]
\centering
\begin{tabular}{|c|c|>{\columncolor[gray]{0.1}}c|c|c|>{\columncolor[gray]{0.1}}c|c|c|}
 \hline
        $m_{\pi}$ & 138 MeV & \ \ & $g_{q\sigma}$ & -1 & \ \ & $g_{\rho p,n}$ & $\pm4.55$ \\ \hline
        $m_{K}$ & 498 MeV &  & $g_{s\zeta}$ & -1 &  & $g_{\rho \Lambda}$ & 0 \\ \hline
        $m_{\omega}$ & 783 MeV &  & $g_{\sigma N}^{(1)}$ & -9.45 &  & $g_{\rho{\Sigma^{\pm}}}$ & $\pm3.63$ \\ \hline
        $m_{\rho}$ & 761 MeV &  & $g_{\sigma \Lambda}^{(1)}$ & -7.62 &  & $g_{\rho {\Xi^{\pm}}}$ & $\pm1.816$ \\ \hline
        $m_{\phi}$ & 1019 MeV &  & $g_{\sigma \Sigma}^{(1)}$ & -5.83 &  & $g_{\phi N}$ & 0 \\ \hline
        $m_{0q}$ & 253 MeV &  & $g_{\sigma \Xi}^{(1)}$ & -4.89 &  & $g_{\phi \Lambda}$ & -3.34 \\ \hline
        $\delta m_q$ & 56 MeV &  & $g_{\sigma B}^{(2)}$ & 3.21 &  & $g_{\phi \Sigma}$ & -3.34 \\ \hline
        $ m_s$ & 130 MeV &  & $g_{\zeta N}^{(1)}$ & -0.899 &  & $g_{\phi \Xi}$ & -6.69 \\ \hline
        $\sigma_0$ & -93.0 MeV &  & $\zeta_0$ & -106.77 MeV &  & $V_0$ & $-(229^4) \mathrm{MeV}^{4}$ \\ \hline        
        $f_{\pi}$ & 93 MeV &  & $g_{\zeta \Lambda}^{(1)}$ & -3.49 &  & $k_0$ & 242$^2$ $\mathrm{MeV}^2$ \\ \hline
        $f_{K}$ & 122 MeV &  & $g_{\zeta \Sigma}^{(1)}$ & -6.02 &  & $k_1$ & 4.818 \\ \hline
        $m_0$ & 759 MeV &  & $g_{\zeta \Xi}^{(1)}$ & -7.35 &  & $k_2$ & 23.3 \\ \hline
        $T_0$ & 180 MeV &  & $g_{\zeta B}^{(2)}$ & 0 &  & $k_4$ & 76$^4$ $\mathrm{MeV}^4$ \\ \hline
        $a_0$ & 3.51 &  & $g_{\omega N}$ & 5.45 &  & $k_6$ & $10^{-4}$ $\mathrm{MeV}^{-2}$ \\ \hline
        $a_1$ & -11.67 &  & $g_{\omega \Lambda}$ & 6 &  & $\beta_2$ & 1500 \\ \hline
        $a_2$ & 9.33 &  & $g_{\omega \Sigma}$ & 8.175 &  & $Z_{\phi}$ & 2.239 \\ \hline
        $b_3$ & -0.53 &  & $g_{\omega \Xi}$ & 4.905 &  & $Z_{\omega}$ & 1.322 \\
\hline

\end{tabular}
\caption{List of default parameters and coupling constants of the CMF model.}\label{param-table}
\end{table}
\bibliography{main}

\end{document}